\documentclass[preprint]{aastex63}
\usepackage{amsmath}
\shorttitle{EXPONENTIAL MODES OF ORTHOGONAL POLARIZATION}
\shortauthors{McKINNON}

\begin{document}

\title{Exponential Fluctuations in the Modes of Orthogonal Polarization in Pulsar Radio Emission}
\author{M. M. McKinnon}
\affiliation{National Radio Astronomy Observatory, Socorro, NM \ 87801\ \ USA}

\begin{abstract}

A statistical model for the polarization of pulsar radio emission is enhanced to account for the 
heavy modulation of the emission, the possible covariance of the Stokes parameters, and the observed 
asymmetries in the distributions of total intensity, polarization, and fractional polarization by 
treating the intensities of the orthogonal polarization modes as exponential random variables. The 
model is used to derive theoretical distributions to compare with what is observed. The resulting 
distributions are unimodal and generally asymmetric. The unimodality arises from the model's 
fundamental assumption that the orthogonal modes are superposed. The asymmetry originates primarily 
from different fluctuations in mode intensities. The distributions of fractional polarization are 
truncated at the degree of linear and circular polarization intrinsic to the modes. A number of 
observable parameters that quantify the statistical properties of the emission and its polarization
are derived and are shown to be functions only of the ratio of the modes' mean intensities, $M$, 
suggesting their spectra coevolve according to the frequency dependence of $M$. This particular 
implementation of the model requires the modes to fluctuate differently in order to replicate the 
observations. Since a single underlying emission mechanism seems unlikely to selectively modulate 
the mode intensities, the different fluctuations are attributed either to different emission 
mechanisms for the modes or to mode-dependent propagation or scattering effects in the pulsar 
magnetosphere.

\end{abstract}

\section{INTRODUCTION}

Single pulse polarization observations of pulsars are made in an attempt to understand their radio
emission mechanism and the propagation of radio waves in their magnetospheres. The observations
reveal the emission is highly linearly polarized and heavily modulated, often switching randomly 
between orthogonally polarized states (Manchester, Taylor, \& Huguenin 1975; Cordes, Rankin, \& 
Backer 1978, hereafter CRB; Stinebring et al. 1984a, hereafter S84a). The orthogonal polarization 
modes (OPMs) can be elliptically polarized with each mode prefering a particular sense of circular 
polarization. The results of these observations have historically been presented as color-coded or 
grey-scale histograms of fractional linear polarization, fractional circular polarization, and 
polarization position angle at multiple locations across a pulsar's pulse (e.g. CRB, S84a). A 
number of factors can affect the appearance of the histograms, such as how the modes interact (e.g. 
do they occur simultaneously or separately?), the statistical character of their fluctuations, the 
degree and ellipticity of their polarization, and the level of instrumental noise. These factors 
must be understood in order for the histograms to be properly interpreted.

McKinnon \& Stinebring (1998, hereafter MS1) developed a statistical model for the OPMs in pulsar 
radio emission. The model provides a general statistical framework for deriving the distributions 
of total intensity, polarization, fractional polarization, and polarization position angle given
the distributions of the mode intensities. They proposed the emission consists of two independent,
completely linearly polarized, simultaneously occurring, orthogonal modes of polarization. The 
model treats the mode intensities as random variables (RVs) to account for the emission's 
variability. Subsequent work accounted for the circular polarization of the modes (McKinnon \&
Stinebring 2000, hereafter MS2; McKinnon 2002, hereafter M02). In their detailed implementation 
of the model, MS1 assumed the mode intensities were Gaussian RVs. While the distributions they 
derived were qualitatively similar to what is observed, the implementation did not address the 
heavy modulation of the observed emission, incorporate the possible covariance of the Stokes 
parameters, or replicate some asymmetries observed in distributions of the Stokes parameters and 
fractional polarization. For example, if the total intensity of the emission follows Gaussian 
statistics, the ratio of its standard deviation to its mean (the modulation index, $\beta$) must 
be $\beta < 0.2-0.3$ for the intensity to be positive definite (M02). However, observed modulation 
indices are often much larger, sometimes exceeding unity (e.g. Bartel, Sieber, \& Wolszczan 1980, 
hereafter BSW). The implementation should accommodate large values of $\beta$ because OPMs tend 
to occur where the emission is heavily modulated (McKinnon 2004, hereafter M04). MS1 also assumed 
the standard deviations of the Gaussian mode intensities were equal thereby stipulating the Stokes 
parameters (specifically I and Q in their analysis) were also independent RVs. This assumption 
simplified the derivation of the distribution of fractional polarization. In general, the Stokes 
parameters are covariant (MS1), and a more robust implementation of the model should provide for 
the correlation. The assumption of Gaussian mode intensities always produces Gaussian, and thus 
symmetric, distributions of the Stokes parameters and Gaussian-like distributions of fractional 
polarization (MS1; M02). While similar distributions are observed, other distributions have 
shapes that are more complex or asymmetric.

The purpose of this paper is to incorporate the heavy modulation of the emission, provide for the 
covariance of the Stokes parameters, and replicate asymmetries in the observed distributions of 
the Stokes parameters and fractional polarization by assuming the mode intensities are exponential, 
instead of Gaussian, RVs in the model's implementation. An exponential RV has the property that its 
mean and standard deviation are equal, and its modulation index is consequently equal to one. The 
assumption of exponential mode intensities also results in the Stokes parameters generally being 
covariant. Asymmetries in distributions of the Stokes parameters and fractional polarization 
arise from the different statistical properties of the exponential mode intensities.

The paper is organized as follows. The general framework of the MS1 statistical model is briefly 
reviewed in \S\ref{sec:model}. The framework is used to derive the distributions of total intensity, 
linear polarization, fractional linear polarization, and polarization position angle when the mode 
intensities are exponential RVs. Parameters are derived to quantify the statistical properties of 
the emission and its polarization. The implementation is then extended to include circular 
polarization and to account for instrumental noise. In \S\ref{sec:simulate}, a numerical simulation 
of the model is used to replicate observed distributions of fractional polarization and position 
angle in PSR B1929+10 and PSR B2020+28. A comparison is made between the results obtained from the 
model's treatment of exponential and Gaussian RVs in \S\ref{sec:discuss}. The implications of the 
analysis for the generation of OPMs and the frequency dependence of the emission and its 
polarization are also discussed.

\section{POLARIZATION STATISTICAL MODEL}
\label{sec:model}

\subsection{Model Overview}

The MS1 statistical model for the polarization of pulsar radio emission is reviewed here for 
convenience and reference for the sections that follow. The simplest form of the model proposes 
that the polarization of the emission is determined by the incoherent superposition of two completely 
linearly polarized, orthogonal modes. The model is consistent with the theoretical proposition 
that OPMs are the natural modes of wave propagation in the pulsar magnetosphere (Melrose 1979;
Onishchenko 1981; Allen \& Melrose 1982; Barnard \& Arons 1986). The model accounts for the 
stochastic nature of OPMs by representing their flux densities by the RVs $X_1$ and $X_2$, which 
are distributed according to the functions, $f_1$ and $f_2$, respectively. The RVs have means 
$\mu_j$ and standard deviations $\sigma_j$. The primary mode is designated as $X_1$ 
($\mu_1 > \mu_2$), and $X_2$ is the secondary mode. Since the modes are incoherent, they propagate 
independently, and the total intensity of the combined emission is the sum of the mode flux 
densities (Chandrasekhar 1960). To simplify the analysis, but without loss of generality, a 
polarization position angle of $\psi=0$ is assigned to the primary mode, and a position angle of 
$\psi=\pi/2$ is assigned to the secondary mode. So defined, the vectors representing the mode 
polarizations are antiparallel to one another and form a diagonal along the Q-axis in the 
Poincar\'e sphere. Since the polarization vectors are antiparallel, the amplitude of the resultant 
polarization vector is the difference between the mode polarization amplitudes, and the tip of the
vector resides in one hemisphere of the Poincar\'e sphere or the other depending upon which mode 
is instantaneously the stronger of the two. With these assumptions and definitions, the Stokes 
parameter I of the combined emission is the sum of the mode intensities, ${\rm I} = X_1 + X_2$, 
and the Stokes parameter Q is their difference, ${\rm Q} = X_1 - X_2$. The remaining Stokes 
parameters are ${\rm U}={\rm V}=0$. The amplitude of the linear polarization is ${\rm L=|Q|}$. 
General expressions for the probability distributions of I, Q, and L can be derived from these 
definitions of the Stokes parameters (MS1). 
 
\begin{equation}
f_{\rm I}(x) = f_1*f_2
\label{eqn:I}
\end{equation}

\begin{equation}
f_{\rm Q}(x) = f_2\star f_1 
\label{eqn:Q}
\end{equation}

\begin{equation}
f_{\rm L}(x) = f_2\star f_1 + f_1\star f_2
\label{eqn:L}
\end{equation}

\noindent The asterisk in equation~\ref{eqn:I} denotes convolution, and the stars in 
equations~\ref{eqn:Q} and~\ref{eqn:L} denote correlation. In this idealized model, the distribution 
of polarization position angle consists of two delta functions separated by $\pi/2$ radians. The 
frequency of occurrence of the primary mode (the amplitude of the larger delta function) is the 
probability that the Stokes parameter Q exceeds zero.

\begin{equation}
\nu_1 = \int_{0}^{\infty}f_{\rm Q}(x)dx
\label{eqn:nu1}
\end{equation}

\noindent Instrumental noise has been ignored in the derivation of equations~\ref{eqn:I}-\ref{eqn:nu1},
but is incorporated in the analysis in \S\ref{sec:noise}.

\subsection{Distributions from Exponential Mode Intensities}
\label{sec:dist}

In their detailed implementation of the polarization model's statistical framework, MS1 assumed the 
mode intensities, $X_1$ and $X_2$, were Gaussian RVs. Their analysis is repeated here, but with the 
mode intensities taken to be exponential RVs. The distribution of an exponential RV is given by

\begin{equation}
f_j(x) = {1\over{\mu_j}}\exp(-x/\mu_j), \qquad x\ge 0,
\label{eqn:fexp}
\end{equation}

\noindent where the subscripts $j=1,2$ denote the primary and secondary modes, respectively. The 
distributions of I and Q can be found by inserting the exponential distributions, $f_1$ and $f_2$, 
from equation~\ref{eqn:fexp} into equations~\ref{eqn:I} and~\ref{eqn:Q}. 

\begin{equation}
f_{\rm I}(x) = {1\over{\mu_1-\mu_2}}[\exp(-x/\mu_1)-\exp(-x/\mu_2)],\qquad x\ge 0
\label{eqn:INE}
\end{equation}

\begin{eqnarray}
f_{\rm Q}(x) & = & {1\over{\mu_1+\mu_2}}\exp(-x/\mu_1), \qquad x\ge 0, \nonumber \\
             &  & {1\over{\mu_1+\mu_2}}\exp(x/\mu_2), \qquad x<0.
\label{eqn:Qexp}
\end{eqnarray}

\noindent Similarly, from equations~\ref{eqn:L} and~\ref{eqn:fexp}, the distribution of linear 
polarization is

\begin{equation}
f_{\rm L}(x) = {1\over{\mu_1+\mu_2}}[\exp(-x/\mu_1)+\exp(-x/\mu_2)],\qquad x\ge 0.
\label{eqn:Lexp}
\end{equation}

\noindent From equations~\ref{eqn:nu1} and~\ref{eqn:Qexp}, the frequency of occurrence of the 
primary mode in the position angle distribution is 

\begin{equation}
\nu_1 = {M\over{M+1}},
\label{eqn:foo}
\end{equation}

\noindent where $M=\mu_1/\mu_2$ is the ratio of the mode mean intensities. The frequency of 
occurrence of the secondary mode is $\nu_2=1-\nu_1 = 1/(M+1)$, and the relative frequency of 
occurrence of the modes is $\nu_1/\nu_2=M$. The mean of Stokes Q normalized by the mean total 
intensity, $m$, is equal to the difference between the frequencies of occurrence of the modes
(i.e. the difference between the amplitudes of the two delta functions comprising the position 
angle distribution).

\begin{equation} 
m={\mu_{\rm Q}\over{\mu_{\rm I}}}={\mu_1-\mu_2\over{\mu_1+\mu_2}}=\nu_1-\nu_2={M-1\over{M+1}}
\label{eqn:mavg}
\end{equation}

The joint probability density of Q and I is needed to derive the distribution of fractional 
linear polarization. The detailed procedure for deriving the joint probability density and 
the distribution of fractional polarization is given in the Appendix. The resulting distribution 
of fractional linear polarization is

\begin{equation}
f_{\rm ml}(z) = {(1-m^2)(1+m^2z^2)\over{(1-m^2z^2)^2}}, \qquad 0\le z\le 1.
\label{eqn:mdist}
\end{equation}

When the mean intensities of the modes are equal, $\mu_1=\mu_2=\mu$, the total intensity 
follows a gamma distribution, the linear polarization is exponentially distributed, the 
fractional linear polarization is uniformly distributed, and the two delta functions in the 
position angle distribution have equal amplitudes ($\nu_1=\nu_2=1/2$).

\begin{equation}
f_{\rm I}(x) = {x\over{\mu^2}}\exp(-x/\mu), \qquad x\ge 0
\label{eqn:IEQ}
\end{equation}

\begin{equation}
f_{\rm L}(x) = {1\over{\mu}}\exp(-x/\mu), \qquad x\ge 0
\end{equation}

\begin{equation}
f_{\rm ml}(z)=1, \quad 0\le z\le 1
\end{equation}

\subsection{Model Statistical Parameters}

A number of parameters can be derived from the definitions of the Stokes parameters and their 
distributions to quantify the statistical properties of the emission and its polarization. The 
modulation index of the total intensity is the ratio of its standard deviation to its mean.

\begin{equation}
\beta = {\sigma_{\rm I}\over{\mu_{\rm_I}}} = 
{(\mu_1^2+\mu_2^2)^{1/2}\over{\mu_1 + \mu_2}} = {(M^2+1)^{1/2}\over{M+1}}
\label{eqn:beta}
\end{equation}

\noindent The Stokes parameters I and Q are generally covariant. The I-Q correlation coefficient
is

\begin{equation}
r_{\rm IQ} = {{\rm Cov(I,Q)}\over{\sigma_{\rm I}\sigma_{\rm Q}}} = 
{\mu_1^2-\mu_2^2\over{\mu_1^2+\mu_2^2}} =  {M^2-1\over{M^2+1}}.
\label{eqn:IQcorr}
\end{equation}

\noindent The mean linear polarization is 

\begin{equation}
\mu_{\rm L} = \int_{0}^{\infty}xf_{\rm L}(x)dx = {\mu_1^2 + \mu_2^2\over{\mu_1 + \mu_2}},
\end{equation}

\noindent and the mean linear polarization normalized by the mean total intensity is

\begin{equation}
\bar{\rm L}={\mu_{\rm L}\over{\mu_{\rm I}}} = {M^2+1\over{(M+1)^2}}.
\end{equation}

\noindent Lastly, the mean of the distribution of fractional linear polarization is 

\begin{equation}
ml = \int_0^1 zf_{\rm ml}(z)dz = 1+{1-m^2\over {2m^2}}\ln{(1-m^2)}, \qquad 0<m<1.
\label{eqn:ml}
\end{equation}

\begin{deluxetable}{cc}
\tablenum{1}
\tablewidth{360pt}
\tablecaption{Model Statistical Parameters}
\tablehead{
  \colhead{Parameter} & \colhead{Expression}}
\startdata
Normalized Mean Q, $m$ & $(M-1)/(M+1)$ \\
 & \\
Normalized Mean L, $\bar{\rm L}$ & $(M^2+1)/(M+1)^2$ \\
 & \\
Primary Mode Frequency of Occurrence, $\nu_1$ & $M/(M+1)$ \\
 & \\
Mean Fractional Linear Polarization, $ml$ & $1+{1-m^2\over{2m^2}}\ln{(1-m^2})$ \\
 & \\
I-Q Correlation Coefficient, $r_{\rm IQ}$ & $(M^2-1)/(M^2+1)$ \\
 & \\
Intensity Modulation Index, $\beta$ & $(M^2+1)^{1/2}/(M+1)$ \\
\enddata
\end{deluxetable}

\begin{figure}[htb!]
\plotone{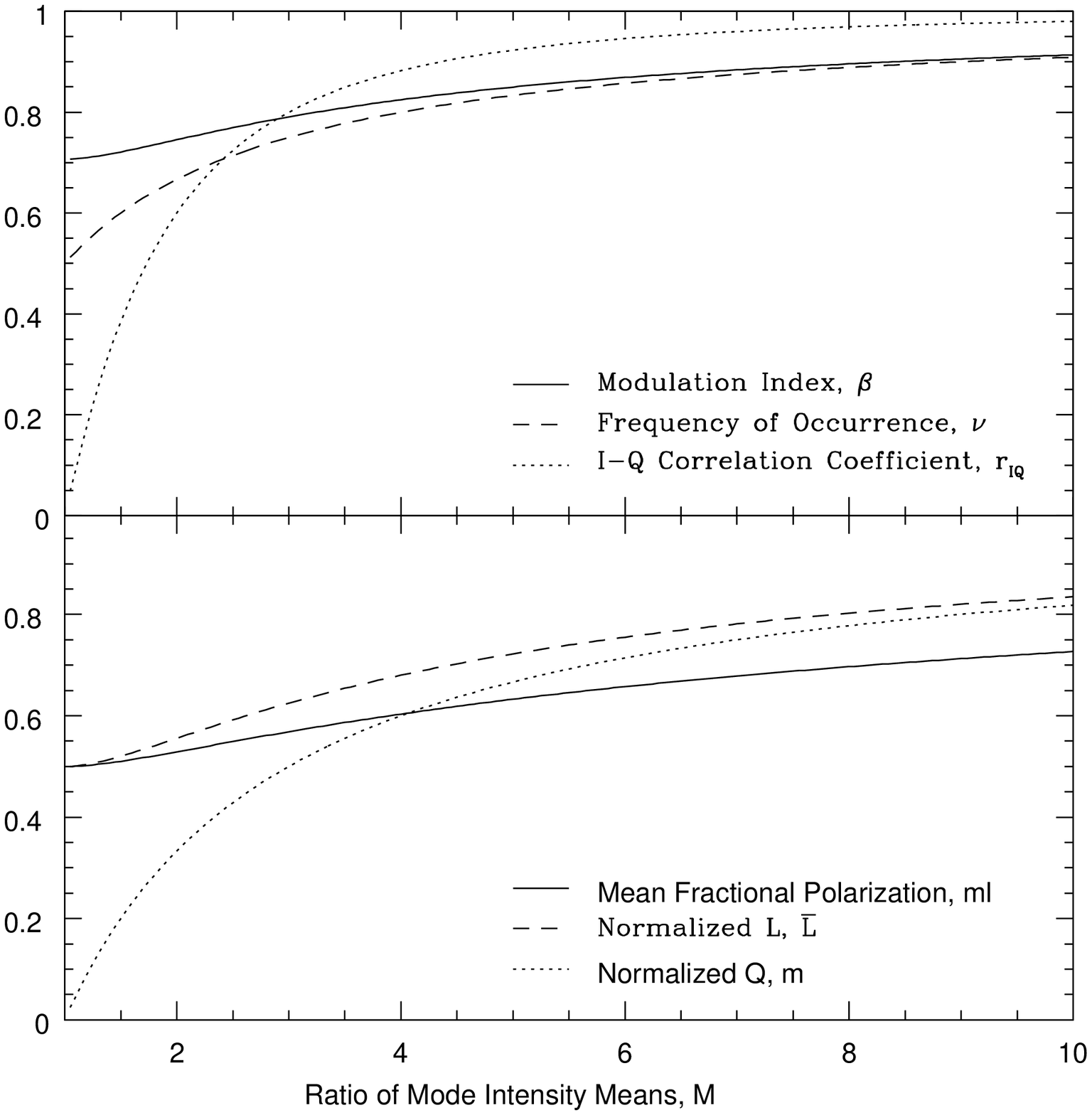}
\caption{Dependence of model statistical parameters upon the ratio of mode mean intensities, $M$.
The top panel shows how the intensity modulation index, primary mode frequency of occurrence, and
correlation coefficient between the Stokes parameters I and Q vary with $M$. The bottom panel shows 
the dependence of the mean of the fractional linear polarization distribution, the normalized
linear polarization, and the normalized Stokes parameter Q upon $M$.}
\label{fig:foo}
\end{figure}

The statistical parameters are summarized in Table 1 and shown in Figure~\ref{fig:foo}.
Interestingly, all of them are functions only of the ratio of the modes' mean intensities, $M$, and 
are independent of the individual values of $\mu_1$ and $\mu_2$. The parameters' sole dependence 
upon $M$ arises because all parameters were derived by taking ratios of various combinations of 
means and standard deviations, and an exponential RV possesses the properties that its distribution
is characterized by a single quantity, $\mu$, and its mean and standard deviation are equal. All of 
the parameters are, in principle, observable quantities. When $M=1$, Figure~\ref{fig:foo} shows the 
modes occur with equal frequency ($\nu_1=1/2$), the Stokes parameters I and Q are not correlated 
($r_{\rm IQ}=0$), the modulation index is $\beta=1/\sqrt{2}$, the normalized mean of Stokes Q is 
$m=0$, and $\bar{\rm L}=ml=1/2$. As $M$ becomes large, all of the parameters approach unity. This 
behavior results from the emission being dominated by the primary mode. In this case, the Stokes 
parameters I and Q are highly correlated, and the emission is highly polarized and heavily modulated. 
The primary mode frequency of occurrence, $\nu_1$, and the intensity modulation index, $\beta$, 
closely track one another at large $M$ with a localized slope varying as $\sim 1/M^2$. The parameters 
$m$ and $\bar{\rm L}$ also track one another, but with a localized slope varying as $\sim 2/M^2$. The 
modulation index remains large for all values of $M$. Close inspection of the table entries reveals 
relationships between the parameters, e.g. $\bar{\rm L}=\beta^2=\nu_1^2+\nu_2^2=m/r_{\rm IQ}$.

\subsection{Including Circular Polarization}
\label{sec:circ}

The statistical model can be expanded to include circular polarization by assuming the orthogonal 
modes are completely elliptically polarized, in which case the Stokes parameters Q, U, and V of 
the combined emission are (M04)

\begin{equation}
{\rm Q} = \sin\theta_o\cos\phi_o(X_1 - X_2),
\end{equation}

\begin{equation}
{\rm U} = \sin\theta_o\sin\phi_o(X_1 - X_2),
\end{equation}

\begin{equation}
{\rm V} = \cos\theta_o(X_1 - X_2).
\end{equation}

\begin{figure}[htb!]
\plotone{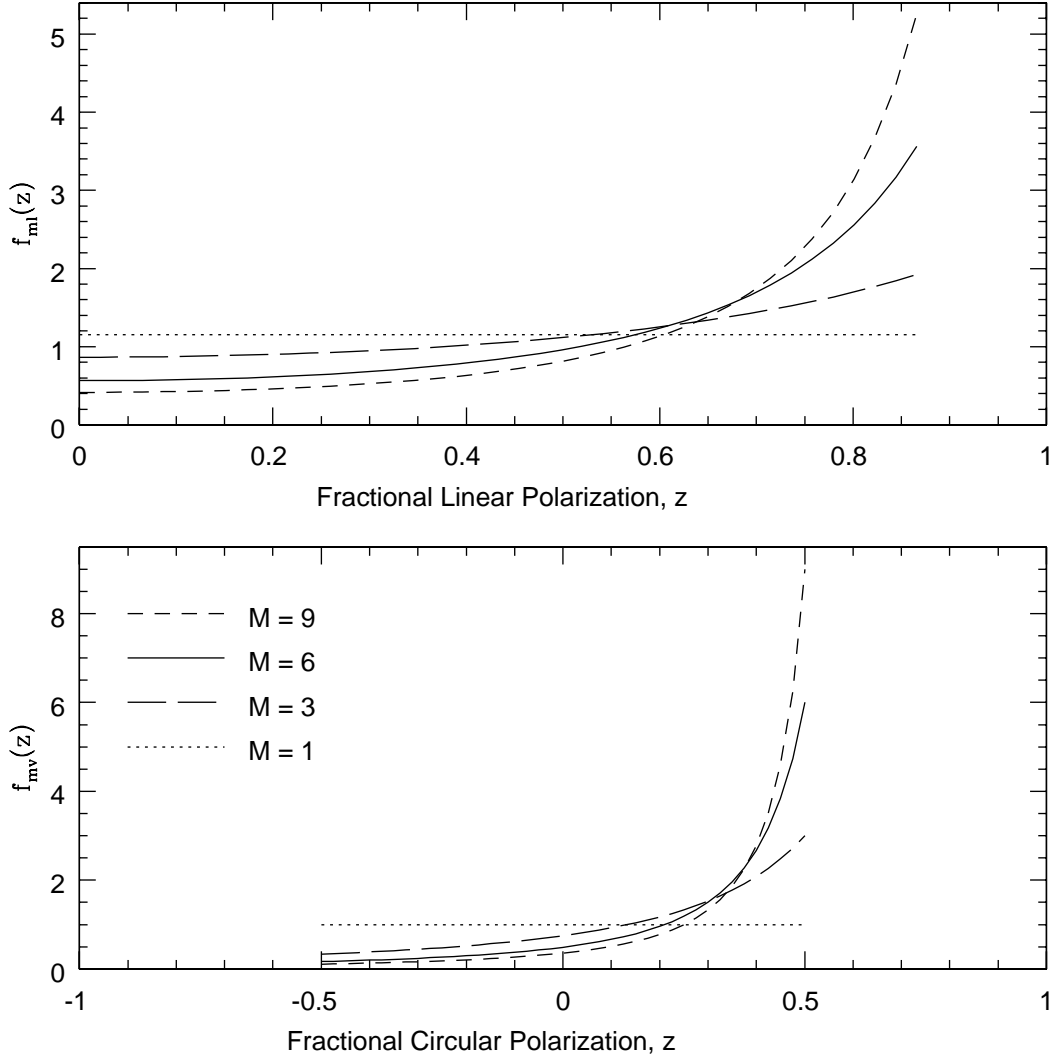}
\caption{Example distributions of fractional linear polarization (top panel) and fractional circular 
polarization (bottom panel) when the mode intensities are exponential random variables. The 
distributions are shown for different values of $M$ when $\theta_o=\pi/3$ radians.}
\label{fig:pol}
\end{figure}

\noindent Here, $\phi_o$ is the azimuth of the primary mode's polarization vector in the 
Poincar\'e sphere and $\theta_o$ is its colatitude. Since each mode is assumed to be completely 
polarized, $\sin\theta_o$ is a mode's degree of linear polarization and $\cos\theta_o$ is its 
degree of circular polarization. The definition of the Stokes parameter I remains unchanged. 
With these definitions, the distributions of Q, U, and V are simply scaled versions of the 
distribution for Stokes Q given by equation~\ref{eqn:Qexp}. To simplify the derivation of 
fractional linear polarization, and again without loss of generality, the azimuth of the 
polarization vector can be assumed to be $\phi_o=0$, so that the linear polarization fluctuations 
are concentrated in Stokes Q, as was done in \S\ref{sec:dist}. Following the analysis described 
in the Appendix, the distributions of fractional linear polarization, $f_{ml}$, and fractional 
circular polarization, $f_{mv}$\footnote{The variable $z$ is used interchangeably to represent 
fractional linear polarization and fractional circular polarization in equation~\ref{eqn:mldist}, 
equation~\ref{eqn:mvdist}, and Figure~\ref{fig:pol}.}, can be shown to be

\begin{equation}
f_{\rm ml}(z) = 
          {1-m^2\over{\sin\theta_o}}{1+(mz/\sin\theta_o)^2\over{[1-(mz/\sin\theta_o)^2]^2}},
          \qquad 0 \le z\le \sin\theta_o
\label{eqn:mldist}
\end{equation}

\begin{equation}
f_{\rm mv}(z) = {1\over{2\cos\theta_o}}{1-m^2\over{[1-(mz/\cos\theta_o)]^2}}
\qquad -\cos\theta_o\le z\le \cos\theta_o
\label{eqn:mvdist}
\end{equation}

\noindent Example distributions of fractional polarization are shown in Figure~\ref{fig:pol} for 
different values of $M$. The distributions are generally skewed towards large values of fractional 
polarization, and particularly so for large values of $M$. The skewness, or asymmetry, of the 
circular distribution can be acute, with values ranging from $1/(2M\cos\theta_o)$ at $-\cos\theta_o$, 
to $M/(2\cos\theta_o)$ at $\cos\theta_o$. The skewness of both distributions is due to the primary 
mode dominating the polarization of the combined emission. The distributions of fractional linear 
and circular polarization are also abruptly truncated at $\sin\theta_o$ and $\pm\cos\theta_o$, 
respectively. The truncations highlight the fact that the fractional linear or circular 
polarization of the combined emission can never exceed the degree of linear or circular polarization 
intrinsic to an individual mode. Consequently, the ranges of the fractional polarization 
distributions are anti-correlated, e.g. the distribution of fractional circular polarization is
narrow when the distribution of fractional linear polarization is wide, and vice versa. 

The mean of the distribution of fractional circular polarization is

\begin{equation}
mv = \cos\theta_o\Biggl[{1\over{m}}
   + {1-m^2\over{2m^2}}\ln\Biggl({1-m\over{1+m}}\Biggr)\Biggr], \qquad 0<m<1.
\end{equation}

\noindent The mean of the distribution of fractional linear polarization is simply the result given 
by equation~\ref{eqn:ml} multiplied by $\sin\theta_o$. The equations for mean fractional 
polarization are not valid for $m=0\ (M=1)$. In that particular case, the distributions of fractional 
polarization are uniform, the mean value of the fractional linear polarization is $ml=\sin\theta_o/2$, 
and the mean value of the fractional circular polarization is $mv=0$.

\subsection{Accounting for Instrumental Noise}
\label{sec:noise}

\begin{figure}
\plotone{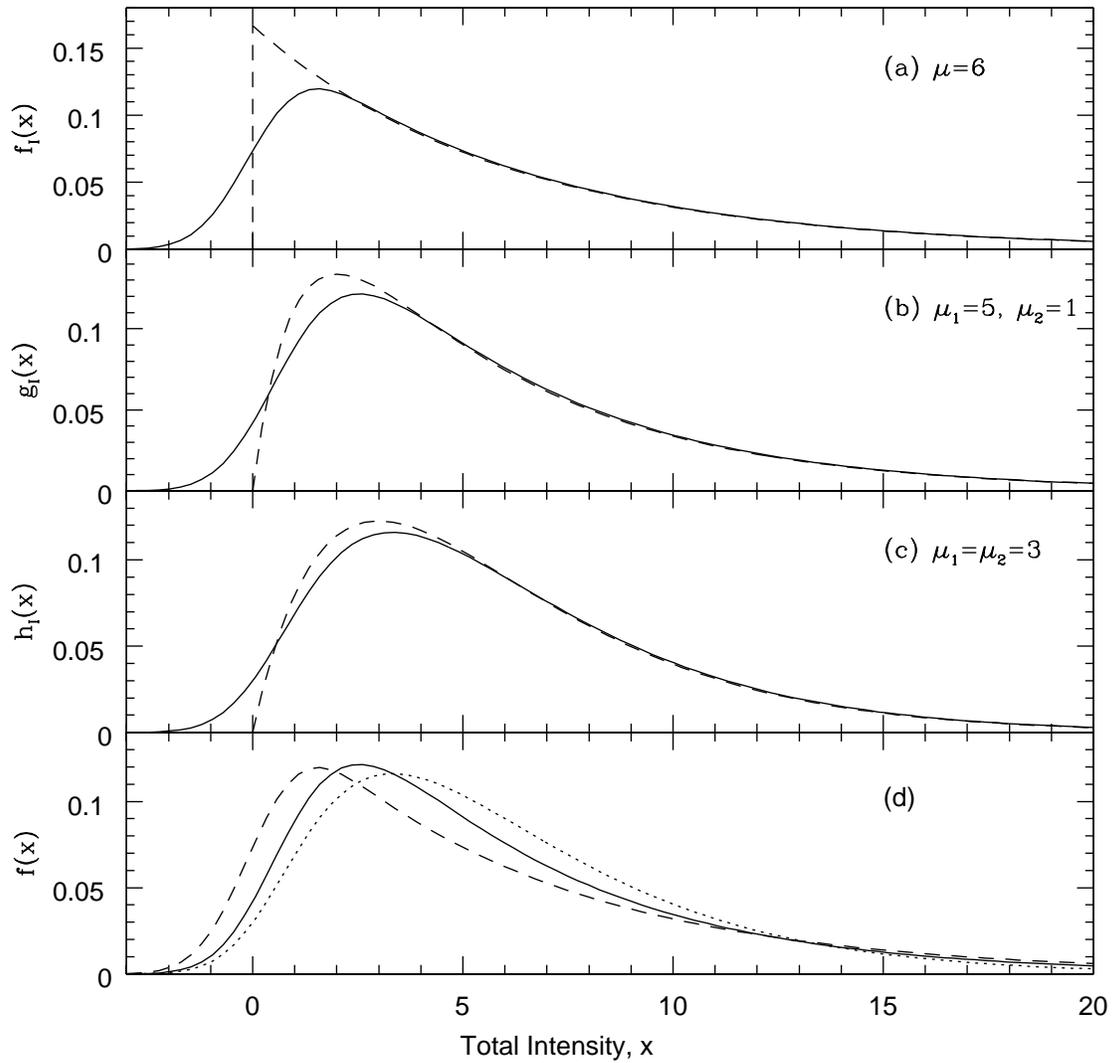}
\caption{Comparisons between total intensity distributions with instrumental noise (solid line) and 
without noise (dashed line). Panel (a) shows the distributions when the total intensity 
is exponentially distributed (equations~\ref{eqn:fexp} \& \ref{eqn:temp} with $\mu=6$). Panel 
(b) shows the distributions when the mode intensities are exponentially distributed, but with 
different means ($\mu_1=5$ and $\mu_2=1$; equations~\ref{eqn:INE} \& \ref{eqn:temp2}). Panel (c) 
shows the distributions when the mode intensities are exponentially distributed with identical 
means ($\mu_1=\mu_2=3$; equations~\ref{eqn:IEQ} \& \ref{eqn:temp3}). Panel (d) compares the three 
distributions that include noise from panels (a) through (c). The value of instrumental noise
used in the figure is $\sigma=1$; therefore, the values on the figure's abscissa may be interpreted 
as a signal-to-noise ratio. All distributions in the figure have the same mean, $\mu_{\rm I}=6$.}
\label{fig:modedist}
\end{figure}

The ideal distributions of the Stokes parameters derived in \S\ref{sec:dist} are those that are
intrinsic to the pulsar, at least within the context of the model. The distributions that are 
actually observed are affected by the additive and statistically independent instrumental noise. 
The effect of the noise can be significant because the narrow bandwidths and short sampling 
intervals used historically in single pulse polarization observations can lead to high 
instrumental noise. The observed distributions of the Stokes parameters are the intrinsic ones 
convolved with a zero-mean Gaussian representing the instrumental noise. The solution to the 
convolution is identical to that for the temporal broadening of a Gaussian-shaped radio pulse due 
to multi-path scattering from a thin screen in the interstellar medium (McKinnon 2014, hereafter 
M14). When the pulsar-intrinsic total intensity is exponentially distributed, the observed 
total intensity distribution is the convolution of equation~\ref{eqn:fexp} with the Gaussian 
instrumental noise (see equation 4 of M14)

\begin{equation}
f_{\rm I}(x,\mu) = {1\over {2\mu}}\exp{\Biggl({\sigma^2\over{2\mu^2}}\Biggr)}
           \exp{\Biggl(-{x \over{\mu}}\Biggr)}
           \Biggl[1+{\rm erf} \Biggl({x-\sigma^2/\mu)\over{\sigma\sqrt{2}}}\Biggr)\Biggr],
\label{eqn:temp}
\end{equation}

\noindent where ${\rm erf}(x)$ is the error function and $\sigma$ is the magnitude of the 
instrumental noise. When the emission is comprised of the two OPMs and $\mu_1\neq\mu_2$, the 
distribution of the observed total intensity is the convolution of equation~\ref{eqn:INE} 
with Gaussian noise (see equation 13 of M14)

\begin{equation}
g_{\rm I}(x) = {\mu_1f_{\rm I}(x,\mu_1) - \mu_2f_{\rm I}(x,\mu_2)\over{\mu_1-\mu_2}},
\label{eqn:temp2}
\end{equation}

\noindent where $f_{\rm I}(x,\mu)$ is given by equation~\ref{eqn:temp}. When $\mu_1=\mu_2=\mu$,
the distribution of the observed total intensity is the convolution of equation~\ref{eqn:IEQ} 
with Gaussian noise (see equation 15 of M14)\footnote{Although equation~\ref{eqn:temp3} appears 
to be different from equation $15$ in M14, both equations produce the same result. 
Equation~\ref{eqn:temp3} is the more concise expression of the two.}.

\begin{equation}
h_{\rm I}(x) = {\sigma\over{\mu^2}}{1\over{\sqrt{2\pi}}}\exp{\Biggl(-{x^2\over{2\sigma^2}}\Biggr)}
     +  {(x-\sigma^2/\mu)\over{\mu}}f_{\rm I}(x,\mu)
\label{eqn:temp3}
\end{equation}

\noindent A comparison between the total intensity distributions with and without instrumental 
noise is shown in Figure~\ref{fig:modedist}. The observed distribution of the Stokes parameter Q 
is the convolution of equation~\ref{eqn:Qexp} with Gaussian noise.

\begin{eqnarray}
f_{\rm Q}(x) & = & {1\over{2(\mu_1+\mu_2)}}\Biggl\{ \exp{\Biggl({\sigma^2\over{2\mu_1^2}}\Biggr)}
           \exp{\Biggl(-{x \over{\mu_1}}\Biggr)}
           \Biggl[1+{\rm erf} \Biggl({x-\sigma^2/\mu_1)\over{\sigma\sqrt{2}}}\Biggr)\Biggr] \nonumber \\
     & + & \exp{\Biggl({\sigma^2\over{2\mu_2^2}}\Biggr)}\exp{\Biggl({x \over{\mu_2}}\Biggr)}
           \Biggl[1-{\rm erf} \Biggl({x+\sigma^2/\mu_2)\over{\sigma\sqrt{2}}}\Biggr)\Biggr]\Biggr\}
\label{eqn:Qdist}
\end{eqnarray}

\begin{figure}
\plotone{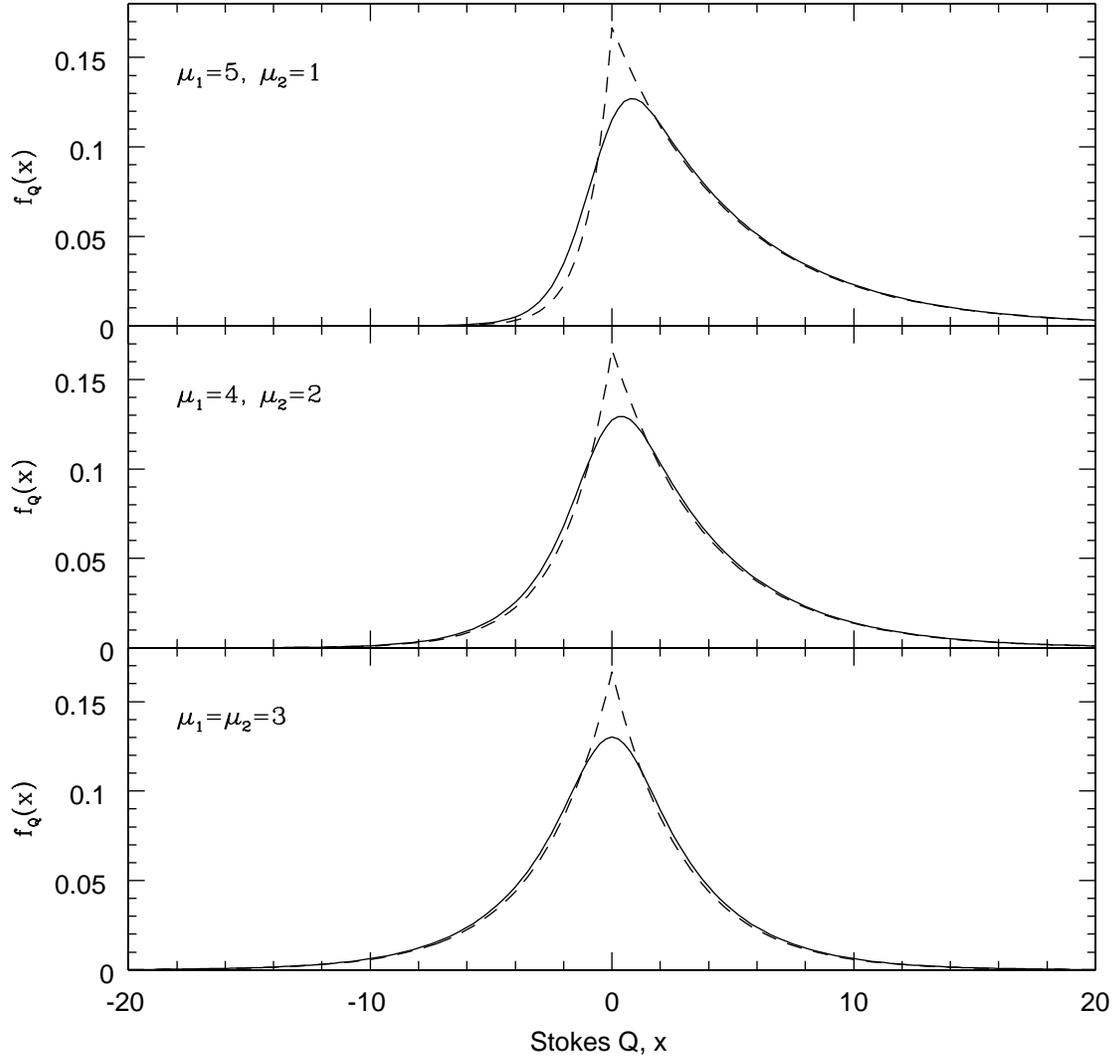}
\caption{Comparisons between distributions of the Stokes parameter Q with instrumental noise (solid 
line) and without noise (dashed line). Each panel compares equation~\ref{eqn:Qexp} to 
equation~\ref{eqn:Qdist} using the noted values of $\mu_1$ and $\mu_2$. The value of instrumental 
noise used in the figure is $\sigma=1$.} 
\label{fig:Qdistfig}
\end{figure}

\noindent A comparison between the distributions of the Stokes parameter Q with and without 
instrumental noise is shown in Figure~\ref{fig:Qdistfig}. If the radio emission is circularly 
polarized, its distribution will be a scaled version of equation~\ref{eqn:Qdist} and what is 
shown in Figure~\ref{fig:Qdistfig}.

The distributions of linear polarization and position angle are derived from the joint probability 
density of the Stokes parameters Q and U (MS1). As the model has been constructed, the pulsar's 
polarization signal is completely contained within the Stokes parameter Q. The Stokes parameter U 
consists of instrumental noise that is independent of Stokes Q. Therefore, the joint probability 
density of Q and U is the product of their individual distributions ($f_{\rm QU}=f_{\rm Q}f_{\rm U}$). 
The distribution of Q is given by equation~\ref{eqn:Qdist}, and the distribution of U is a zero-mean 
Gaussian with noise $\sigma$. After converting the Cartesian coordinates of Q and U in the joint 
density to the polar coordinates of radius and azimuth, the azimuth distribution, $f_{\phi}(\phi)$,  
is found by integrating the joint density over radius.  The linear polarization distribution, 
$f_{\rm L}(r)$, is found by integrating the joint density over azimuth. The joint probability 
density is

\begin{eqnarray}
f(r,\phi) & = & {1\over{2(\mu_1+\mu_2)}}{r\over{\sigma\sqrt{2\pi}}} \nonumber \\
     & \times & \Biggl\{\exp{\Biggl\{-{r^2\over{2\sigma^2}}\Biggl[1-\Biggl(\cos(\phi)
     - {\sigma^2\over{\mu_1 r}}\Biggr)^2\Biggr]\Biggr\}}
     \Biggl\{1+{\rm erf}\Biggl[{r(\cos(\phi)-\sigma^2/\mu_1r)\over{\sigma\sqrt{2}}}\Biggr]\Biggr\} 
     \nonumber \\
     & + & \exp{\Biggl\{-{r^2\over{2\sigma^2}}\Biggl[1-\Biggl(\cos(\phi)
     + {\sigma^2\over{\mu_2 r}}\Biggr)^2\Biggr]\Biggr\}}
     \Biggl\{1-{\rm erf}\Biggl[{r(\cos(\phi)+\sigma^2/\mu_2r)\over{\sigma\sqrt{2}}}\Biggr]\Biggr\} 
     \Biggr\}.
\label{eqn:jointQU}
\end{eqnarray}

\noindent Since azimuth and position angle are related by $\phi=2\psi$, the position angle distribution 
is $f_{\psi}(\psi)=2f_{\phi}[2(\psi-\psi_o)]$, where $\psi_o$ is the position angle of the primary mode
polarization vector. The joint density as written in equation~\ref{eqn:jointQU} places the mode 
peaks at $\phi=0,\pi$ in the azimuth distribution ($\psi=0,\pi/2$ in the position angle distribution). 
The primary mode peak can be arbitarily relocated by selecting a value of $\psi_o$, thereby rotating
the shape of the distribution in angle.

\begin{figure}[htb!]
\plotone{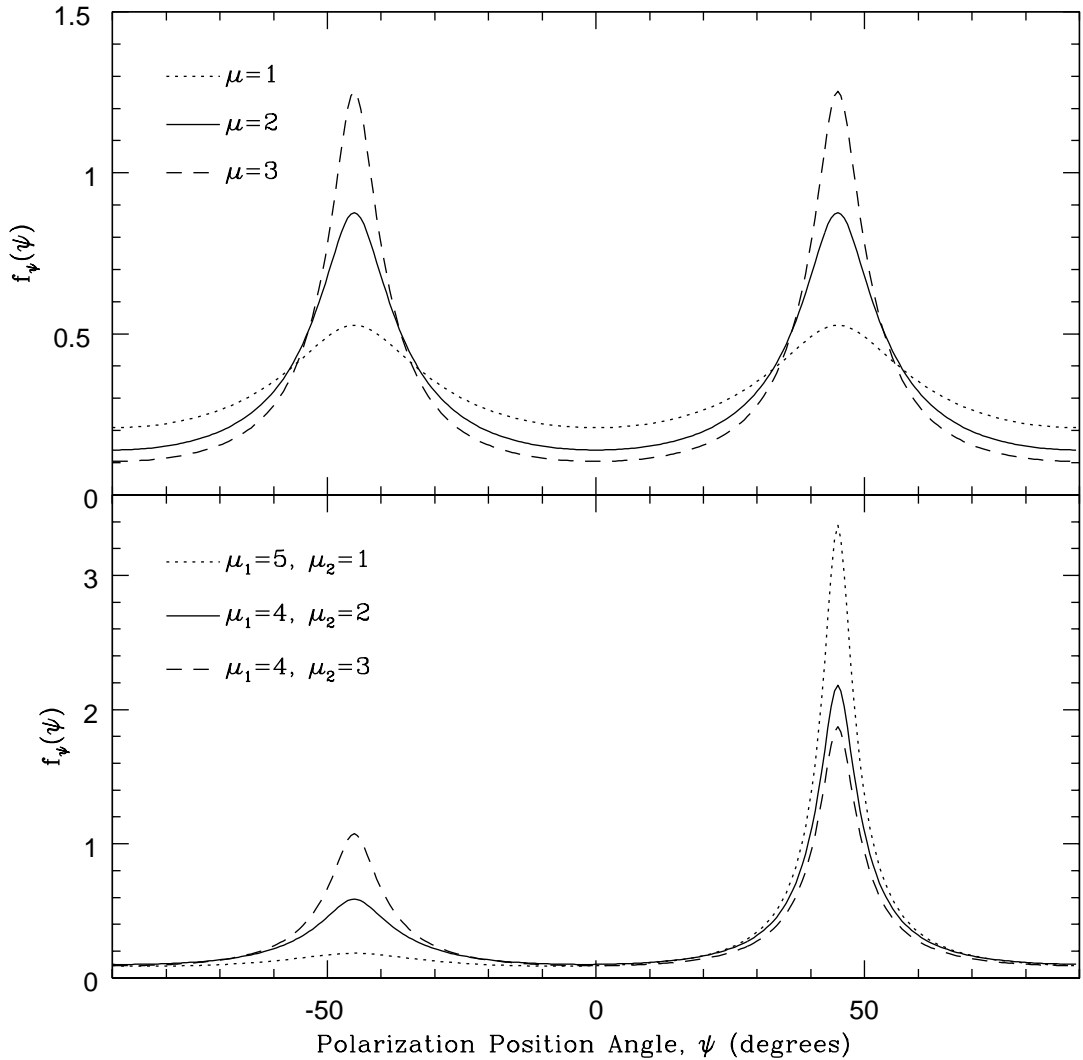}
\caption{Example distributions of polarization position angle for exponentially-distributed mode
intensities. In the top panel, the mean intensities of the modes are equal; their values are 
annotated in the panel. In the bottom panel, the mode mean intensities are not equal; their values
are also listed in the panel. For presentation purposes, the position angle distribution has been 
rotated to place the primary mode peak at $\psi_o=\pi/4$. The value of instrumental noise used in the 
figure is $\sigma=1$.} 
\label{fig:psidist}
\end{figure}

The integration of equation~\ref{eqn:jointQU} was performed numerically for different values of 
$\mu_1$ and $\mu_2$ to generate the example distributions of position angle shown in 
Figure~\ref{fig:psidist}. The top panel of the figure shows example distributions when the mode mean 
intensities are equal. The two peaks representing the modes in the distributions are separated by 
$\pi/2$ radians and occur with equal frequency. The amplitudes of the peaks increase and their widths 
narrow as the value of $\mu$ increases. The ``noise floor'' across the distribution is suppressed as 
$\mu$ increases. The bottom panel of the figure shows the distributions when the mode mean 
intensities are not equal. The amplitude of the primary mode peak exceeds that of the secondary mode. 
The difference between the peak amplitudes increases as $\mu_1$ becomes larger with respect to 
$\mu_2$. The distributions in both panels of the figure resemble the analytical result derived for
Gaussian RVs (see Figure 2 of MS1).

\begin{figure}[htb!]
\plotone{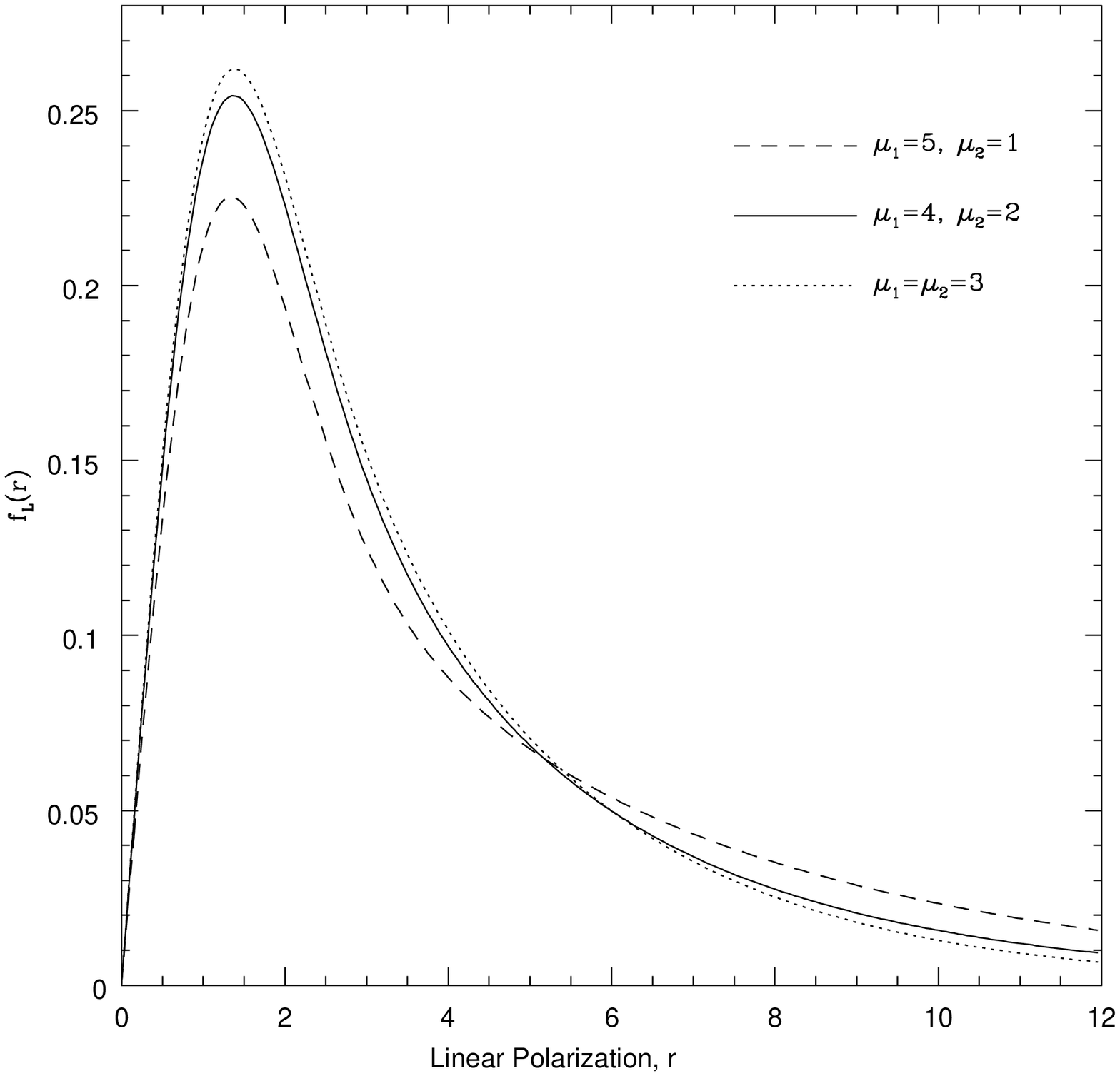}
\caption{Example distributions of linear polarization for exponentially-distributed mode intensities. 
The values of the mode mean intensities are listed in the figure. The value of instrumental noise 
used in the figure is $\sigma=1$.} 
\label{fig:lindist}
\end{figure}

The overall shapes of the resultant distributions of linear polarization are generally similar for 
different values of mode mean intensities (Figure~\ref{fig:lindist}). They resemble a Rician 
distribution, but with highly extended tails. The tail extends further with increasing $M$. The 
resemblance to a Rician distribution arises from the $r\exp{(-r^2)}$ dependence of the joint 
probability density shown in equation~\ref{eqn:jointQU}. For a comparison with the distribution of 
instrumental noise, a pure Rician (Rayleigh) distribution representing the noise contains most data 
samples within $r=3$ and peaks at $r=1$, where $f_{\rm L}(r)=0.607$ is off scale in the figure.

\section{POLARIZATION SIMULATIONS}
\label{sec:simulate}

\begin{figure}[htb!]
\plotone{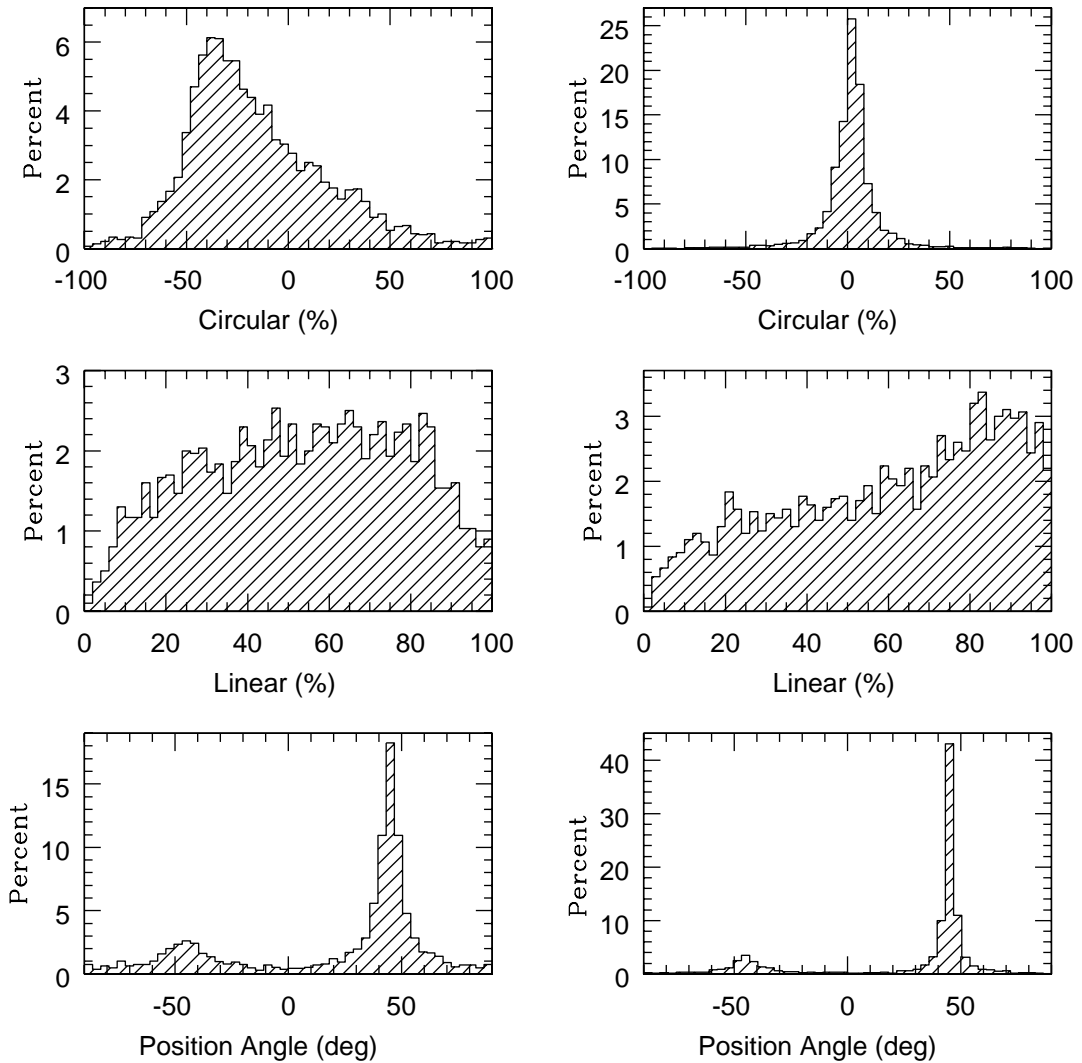}
\caption{Simulated histograms of percentage circular and linear polarization and position angle. 
The left column of panels attempts to replicate distributions observed in PSR B2020+28 at 800 MHz 
(S84b). The parameters used in the simulation are $\mu_1=6$, $\mu_2=2$, $\theta_o=117^\circ$, and
$\sigma=1$. The right column of panels attempts to replicate distributions in PSR B1929+10 at 
1404 MHz (S84a). The parameters used in the simulation are $\mu_1=15.9$, $\mu_2=3.7$, 
$\theta_o=88^\circ$, and $\sigma=1$.}
\label{fig:hist6}
\end{figure}

Numerical simulations were made with the statistical model to replicate observed histograms of 
position angle and fractional circular polarization in PSR B1929+10 at 1404 MHz (Figures 23 and
24, respectively, on page 266 of S84a) and the fractional circular polarization in PSR B2020+28 at 
800 MHz (Figure 13 on page 286 of Stinebring et al. 1984b, hereafter S84b). The modes were modelled 
as exponential RVs. An independent Gaussian noise component with $\sigma=1$ was generated for each 
Stokes parameter. The Stokes parameters were calculated from the equations given in \S\ref{sec:circ}.
The arbitrary value of the primary mode's position angle was set to $\pi/4$, primarily for display 
purposes. The remaining free parameters in the simulation are $\mu_1$, $\mu_2$, and $\theta_o$. The 
simulation generated 3,000 samples of each Stokes parameter. Values of fractional linear polarization, 
fractional circular polarization, and position angle were then calculated from the Stokes parameters 
and binned into histograms of 50 bins each. No detection thresholds were placed on the total 
intensity or linear polarization prior to constructing the histograms. The results of the 
simulations are shown in Figure~\ref{fig:hist6}.

The data used to construct the observed histograms of PSR B2020+28 were recorded near the peak
of the trailing component in its pulse (S84b), where OPMs are obvious and the emission is heavily 
modulated ($\beta\simeq 1.1-1.3$; see Figure 2 of M04). The pulsar is well known for the correlation
between the handedness of its circular polarization and position angle at this location within its
pulse (CRB; S84a). S84b noted the histogram of fractional circular polarization (Figure 13 therein) 
is highly asymmetric with a peak near $z=-0.3$. More specifically, the histogram has a sharp edge at 
about $z=-0.4$ with a tail extending well into positive values of fractional circular polarization. 
They suggested the histogram could be composed of two offset Gaussian components, with each produced 
by one of the modes. They correctly noted that if this were the case, the mode polarizations would 
not be orthogonal. A different interpretation of the histogram is it arises from the superposition 
of elliptically polarized orthogonal modes, and the edge of the histogram is set by the fractional 
circular polarization intrinsic to the modes, as shown by equation~\ref{eqn:mvdist} and 
Figure~\ref{fig:pol}. The simulation of PSR B2020+28 is shown in the three panels on the left side 
of Figure~\ref{fig:hist6}. The model parameters used in the simulation are $\mu_1=6$, $\mu_2=2$, 
and $\theta_o=117^\circ$. The simulated histogram of fractional circular polarization generally 
mimics what is observed, suggesting that non-orthogonal elliptical modes are not required to 
explain the observation. The value of $\theta_o$ used in the simulation indicates the degree of 
circular polarization intrinsic to the primary mode is $\cos(117^\circ)=-0.45$, which is consistent 
with the prevalence of strong circular polarization observed in the pulsar. The observed histogram 
of fractional linear polarization in Figure 12 of S84b extends to $z\simeq 0.8$, while the simulated 
histogram shows a drop off starting at about $z=0.85$. The drop off is caused by the limit on the 
fractional linear polarization imposed by the modes at $\sin(117^\circ)=0.89$. The simulated 
histogram of position angle appears to be a reasonable approximation to what is shown in Figure 12 
of S84b.

The data used to construct the observed histograms of PSR B1929+10 were recorded on the trailing 
edge of the pulse peak (S84a), where OPMs are obvious and the emission is heavily modulated 
($\beta\simeq 0.6-0.7$; see Figure 2 of M04). The peaks in the observed position angle histogram 
(Figure 23 of S84a) were the narrowest of the pulsars S84a observed, with the primary mode 
accounting for almost 44 percent of the data samples recorded. The clear definition of the position 
angle peaks and the large amplitude of the primary mode peak suggest a significant difference 
between the mode means. Despite the high linear polarization implied by the position angle histogram, 
Figure 22 on page 267 of S84a shows the fractional linear polarization of the pulsar at this location 
spans the entire range of possible values ($0\le z\le 1$). The observed histogram of fractional 
circular polarization (Figure 24 therein) is very narrow with a large peak at zero and a slight 
asymmetry composed of a steep edge at positive polarization and a subtle tail at negative 
polarization. The occurence of completely polarized samples in the observed histogram of fractional 
linear polarization and the narrow width of the observed histogram of fractional circular 
polarization suggest the modes are highly linearly polarized, such that the colatitude of the modes' 
polarization vectors is $\theta_o\simeq\pi/2$ (M02). The simulation of PSR B1929+10 is shown in the 
three panels on the right side of Figure~\ref{fig:hist6}. The model parameters used in the simulation 
are $\mu_1=15.9$, $\mu_2=3.7$, and $\theta_o=88^\circ$. The simulated position angle histogram is 
very similar to what is observed. The simulated histogram of fractional linear polarization spans 
its entire range with more samples at high linear polarization, just as equation~\ref{eqn:mldist} 
predicts when $M$ is large and $\theta_o\simeq\pi/2$. The simulated histogram of fractional 
circular polarization is narrow with a large peak and slight asymmetry as predicted by 
equation~\ref{eqn:mvdist}. However, the simulated asymmetry is not quite as discernible as what is 
observed. 

\section{DISCUSSION}
\label{sec:discuss}

\subsection{Comparison between Gaussian and Exponential Intensity Fluctuations}

The distributions of the Stokes parameters I, Q, U, and V, linear polarization, fractional 
polarization, and position angle derived from Gaussian and exponential mode intensities have 
similarities and differences. Overall, the distributions of the Stokes parameters and fractional 
polarization are unimodal. This feature of the distributions is not a coincidence. It is a direct 
result of the statistical model's fundamental assumption that OPMs are superposed (M02). If the 
modes were disjoint, the resulting distributions would generally be bimodal because disjoint modes 
are mutually exclusive, i.e. when one mode is present, the other one is always absent (MS2; M02). 
OPMs could be fundamentally disjoint, but the switching timescale between them remains unresolved 
(van Straten \& Tiburzi 2017).

\begin{deluxetable}{ccc}
\tablenum{2}
\tablewidth{360pt}
\tablecaption{Comparison of Properties of Position Angle Distributions}
\tablehead{
  \colhead{Mode Fluctuations} & \colhead{Exponential} & \colhead{Gaussian}}
\startdata
    & & \\
Difference in Peak Amplitudes & $\sqrt{2/\pi}\mu_{\rm Q}/\sigma$
    & $\sqrt{2/\pi}\mu_{\rm Q}/\sigma$ \\
    & & \\
Primary Mode Peak Width, FWHM & $\ln(2)\sigma/\mu_1$ & 
    $\arcsin(1/\rho)\exp{[-\mu_Q/(k_1\sigma_Q)]}$ \\
    & & \\
Secondary Mode Peak Width, FWHM & $\ln(2)\sigma/\mu_2$ & 
    $\arcsin(1/\rho)\exp{[\mu_Q/(k_2(\rho)\sigma_Q)]}$ \\
    & & \\
\enddata
\end{deluxetable}

\begin{figure}[htb!]
\plotone{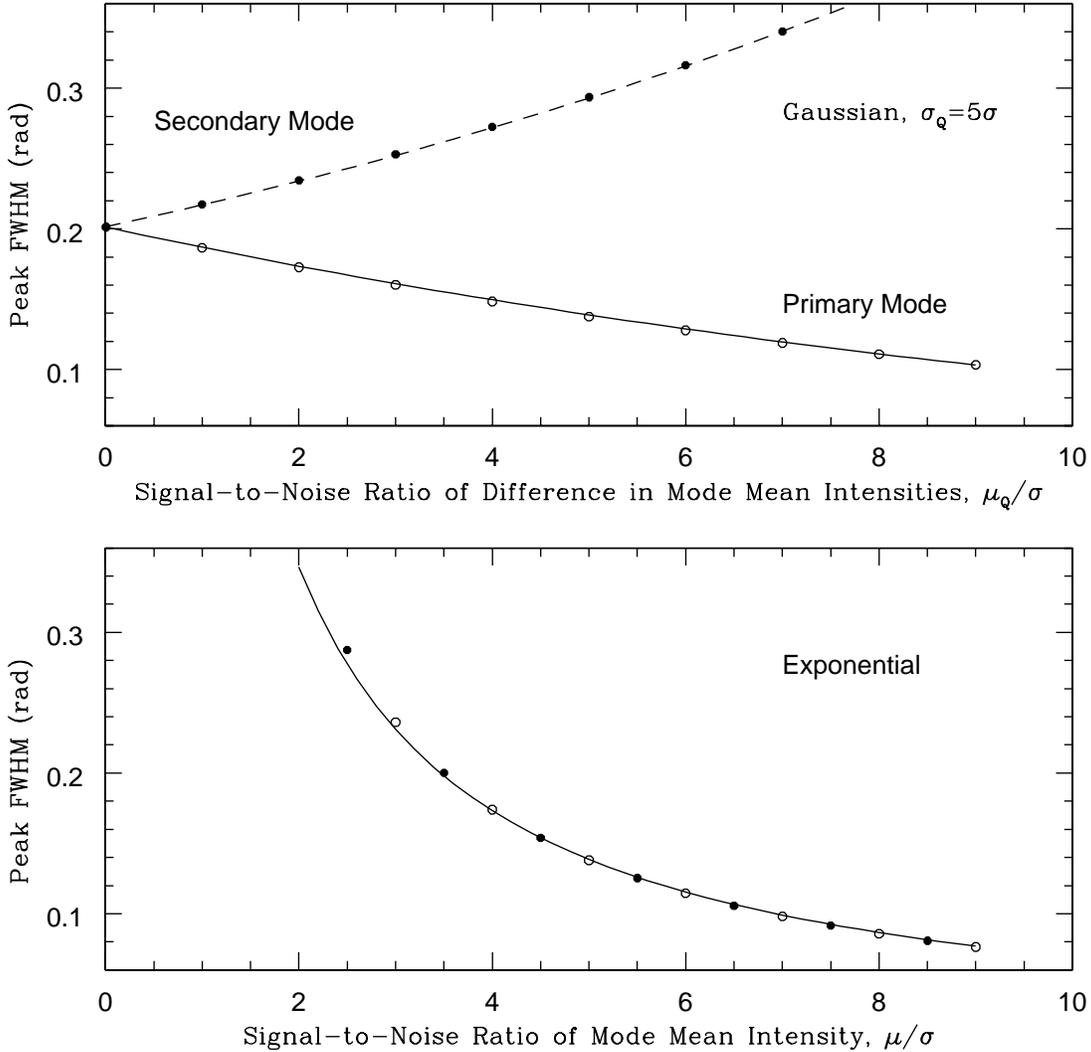}
\caption{Widths of the mode peaks in the position angle distributions for Gaussian fluctuations
(top panel) and exponential fluctuations (bottom panel) in mode intensities. For exponential 
fluctuations, the widths of both mode peaks follow the same track in varying with $\mu_j$. For 
Gaussian fluctuations, the widths of the primary mode peak (solid line) and secondary mode peak 
(dashed line) follow separate tracks in varying with $\mu_{\rm Q}$. The data points in both panels 
represent the peak widths measured from the distributions. Open and closed circles denote the peak 
widths of the primary and secondary modes, respectively. The lines drawn in the panels are 
approximations to the data. Their analytical form is shown in Table 2. The constant values used 
in the figure are $\sigma_{\rm Q}=5\sigma$, $\sigma=1$, $k_1=2.692$, and $k_2(5)=2.666$ (see 
Table 2).} 
\label{fig:width}
\end{figure}

When the mode intensities are exponential RVs, the distribution of the observed Stokes parameter I 
resembles a smoothed, one-sided exponential, and distributions of the observed Stokes parameters Q, 
U, and V resemble smoothed, two-sided exponentials. These distributions are unimodal and generally 
{\it asymmetric}. The modulation index of the total intensity lies in the range $0.71\le\beta\le 1$, 
comparable to what is commonly observed in pulsars (BSW). The distribution of linear polarization 
{\it always} resembles a Rician with an extended tail. The distribution flattens and the tail
becomes longer as $M$ increases. The distributions of fractional polarization are uniform when
$M=1$ and gradually skew upwards towards high values of fractional polarization as $M$ increases.
The distribution of fractional circular polarization can become highly asymmetric when $M$ is large. 
The distributions of fractional linear and fractional circular polarization are abruptly truncated 
at $\sin\theta_o$ and $\pm\cos\theta_o$, respectively, because the fractional polarization of the 
combined emission can never exceed the degree of linear or circular polarization intrinsic to the 
individual modes. Instrumental noise smooths the cutoff, causing the distributions to peak near 
$\sin\theta_o$ and $\cos\theta_o$. Accordingly, the extent of the fractional polarization in 
observed histograms can be used to constrain the ellipticity of the polarization intrinsic to 
the modes. The distribution of polarization position angle is generally bimodal, consisting of two 
peaks sitting atop a uniform plateau and separated by $\pi/2$ radians (Figure~\ref{fig:psidist}). 
As shown in Table 2, the full width at half maximum (FWHM) of the peaks varies as 
$\Delta\psi_j\simeq \ln(2)\sigma/\mu_j$. The functional forms of the peak widths listed in 
the table were found by calculating the widths directly from the position angle distributions and 
approximating them with an equation that best represented the calculated values (see 
Figure~\ref{fig:width}). The FWHM of one mode peak is independent of the other mode. The secondary 
mode peak is generally wider than the primary mode peak, because $\mu_2\le\mu_1$ by definition. 
The difference in the amplitudes of the peaks is directly proportional to the difference in mode 
mean intensities, $\mu_{\rm Q}=\mu_1-\mu_2$ (Table 2). 

When the mode intensities are Gaussian RVs, the Stokes parameters are also Gaussian RVs; therefore, 
their distributions are unimodal and {\it symmetric}. The modulation index of the total intensity 
lies in the range $\beta\le 0.2-0.3$. The distributions of linear polarization and fractional 
polarization resemble a Rician with an extended tail when $\mu_{\rm Q}=0$ ($M=1$; see equation 34 
of McKinnon 2003), but evolve into single, isolated Gaussians as $M$ increases (see MS1, Figure 1 
and equations 12 and 16). The width of the distribution of fractional circular polarization varies 
roughly as the product of the emission's modulation index and the modes' degree of circular 
polarization, $\beta\cos\theta_o$ (see equation 21 of M02). The position angle distributions derived 
from Gaussian RVs are similar to those derived for exponential RVs (compare Figure~\ref{fig:psidist} 
with Figure 2 of MS1). Despite the differences in the statistical character of their fluctuations, 
the difference between mode peak amplitudes is the same for exponential and Gaussian mode intensities 
(Table 2). The FWHM of the Gaussian mode peaks are dependent upon one another through $\mu_{\rm Q}$ 
and $\sigma_{\rm Q}$. When $\mu_{\rm Q}=0$, both the amplitudes and widths of the two peaks are equal, 
with the FWHM of the peaks varying as $\Delta\psi=\arcsin(\rho^{-1})\simeq\rho^{-1}$, where 
$\rho=\sigma_{\rm Q}/\sigma$ (see equation 38 of MS1). As $\mu_{\rm Q}$ increases, the width of the 
primary mode peak narrows exponentially, and the width of the secondary mode peak broadens 
exponentially (Table 2 and Figure~\ref{fig:width}). Thus, the secondary mode peak is generally wider 
than the primary mode peak. When $\mu_{\rm Q}$ exceeds $\sigma_{\rm Q}$, the width of the secondary 
mode peak becomes large, eventually merging into the distribution's noise floor, at which point the 
distribution features only the primary mode peak. The parameters $k_1$ and $k_2(\rho)$ used in the 
expressions for the FWHM of the Gaussian mode peaks in Table 2 were determined for multiple values 
of $\rho$. The value of $k_1$ varied little with $\rho$, and is statistically consistent with a 
constant. The value of $k_2(\rho)$ increases with $\rho$. The values used in the fits to the FWHM 
data points in Figure~\ref{fig:width} are $k_1=2.692$ and $k_2(5)=2.666$.

\subsection{Spectra of the Statistical Parameters}

The analysis of exponential mode intensities was used to derive parameters that quantify the
statistical properties of the emission and its polarization. The analysis shows the parameters
are dependent only on the mode intensity ratio, $M$, and are independent of the mean intensities 
of the individual modes. If the statistical model is a reasonable representation of the emission
and if the fluctuations in mode intensities retain their exponential character over a wide range 
of frequency, the spectral behavior of the parameters should coevolve in frequency in accordance 
with the frequency dependence of $M$ and their dependence upon $M$. For example, both the 
modulation index, $\beta$, and the fractional linear polarization, $\bar{\rm L}$, of many pulsars 
are known to decrease with increasing frequency (BSW; Manchester, Taylor, \& Huguenin 1973; 
Morris, Graham, \& Seiber 1981). This behavior is qualitatively consistent with their dependence 
upon $M$, as listed in Table 1, and with $M$ decreasing with increasing frequency. Similarly, 
Karastergiou et al. (2005) suggested the frequency dependence of a pulsar's fractional linear 
polarization depends upon the total intensity spectra of the two orthogonal modes, or alternatively 
the spectrum of the mode intensity ratio, provided the modes are completely polarized, as assumed 
in this analysis. As the mode intensities become comparable, the radiation will depolarize and the 
modes will occur with nearly equal frequency. Their simultaneous, dual-frequency observations 
showed this to be the case for PSR B1133+16 (Karastergiou et al. 2002).

\subsection{Origin of the Mode Intensity Fluctuations}

The different types of mode intensity fluctuations evaluated within the context of the model's 
statistical framework suggest different interpretations for their physical origins. When MS1 
evaluated the mode intensities as Gaussian RVs, the mode means were generally different, but their 
standard deviations were the same. Within the context of pulsar radio emission and OPMs, this 
scenario could arise from the birefringence of the pulsar's magnetospheric plasma (Melrose 1979; 
Allen \& Melrose 1982). The extraordinary (X) mode propagates as if in vacuum along a straight 
ray path, so its trajectory is largely unaffected by the plasma. The ordinary (O) mode has a 
different index of refraction, and is ducted along the pulsar's open magnetic field lines 
(Onishchenko 1981; Barnard \& Arons 1986). The birefringence of the plasma causes the modes to 
separate spatially, possibly resulting in the difference between their mean intensities 
($\mu_1\neq\mu_2$) at specific locations within the pulse. The fluctuations in mode intensities 
could then be attributed to the underlying emission mechanism, where both modes would be modulated 
in a similar, but independent, way leading to $\sigma_1=\sigma_2$. 

The treatment of mode intensities as exponential RVs in this work requires the mode intensity 
fluctuations to be different ($\sigma_1\neq\sigma_2$) to explain what is observed. It seems difficult 
for a single emission mechanism to modulate the mode intensities differently. One possibility for the 
difference in mode intensity fluctuations is the modes are produced by separate emission mechanisms, 
such as the generation of the X-mode by curvature radiation and the generation of the O-mode by the
acceleration of charged particles along magnetic field lines, as proposed by Cheng \& Ruderman (1979). 
Alternatively, the difference in mode fluctuations could arise from their different propagation or
scattering properties. In this scenario, the intensities of both the X- and O-mode are fundamentally 
modulated by the underlying emission mechanism, but the O-mode intensity fluctuations are altered, 
for example, by stimulated scattering in the magnetosphere where an electron's gyrofrequency is
very large in comparison to the frequency of a photon incident upon it (Blandford \& Scharlemann 1976; 
Sincell \& Krolik 1992; Lyubarskii \& Petrova 1996). Scattering of the X-mode is comparatively 
negligible, so its fluctuations would be largely unaffected. Summarizing, differential refraction 
alone is not sufficient to explain the observed behavior of the orthogonal modes; their intensities 
must also fluctuate differently.

\section{CONCLUSIONS}
\label{sec:conclude}

Distributions of the Stokes parameters and fractional polarization of pulsar radio emission were
derived from a statistical model assuming the intensities of the orthogonal modes comprising 
the emission are exponential random variables. The analysis incorporates both linear and circular 
polarization and accounts for instrumental noise. The resulting distributions of the Stokes 
parameters and fractional polarization are unimodal, as found when the mode intensities follow 
Gaussian statistics. This general feature of the distributions arises from the fundamental 
assumption that the modes are superposed, and is generally consistent with what is observed. 
However, unlike the symmetric distributions produced from Gaussian mode intensities, the 
distributions produced by exponential mode intensities are generally asymmetric. Distributions 
of the emission's fractional polarization are truncated at the degree of linear and circular 
polarization intrinsic to the modes. The asymmetric distributions have been observed in single 
pulse polarization observations of pulsars. Additionally, exponential mode statistics can
replicate the heavy modulation of the intensity and polarization observed in the emission,
whereas Gaussian statistics cannot. The position angle distributions derived from exponential
and Gaussian mode fluctuations are similar in appearance. The width of an exponential mode peak
in the distribution is independent of the other mode, while the widths of the Gaussian mode
peaks are dependent upon one another. The analysis also shows that a number of observables, such 
as modulation index, mode frequency of occurrence, and mean fractional polarization, are functions 
only of the ratio of mode mean intensities, $M$, suggesting their spectral evolution is determined 
primarily by the frequency dependence of $M$. If the statistical model with its implementation of 
exponential mode intensities is a plausible explanation for the polarization of pulsar radio 
emission, some type of mechanism must be responsible for causing different fluctuations in the 
intensities of the orthogonal modes. Since fluctuations produced by a single emission mechanism 
would presumably affect both modes similarly, the difference in mode fluctuations could arise 
from different emission mechanisms for the modes or from mode-dependent propagation or scattering 
effects within the pulsar magnetosphere.

\appendix

\section{JOINT PROBABILITY DENSITY AND FRACTIONAL POLARIZATION}

The joint probability density of Q and I must be found to derive the distribution of fractional 
linear polarization. In the derivation of MS1, Q and I were constructed to be independent, which is 
appropriate for the Gaussian RVs used in their analysis, but not for the exponential RVs considered 
here. From Ross (1984), the joint probability density of two functions, $y_1$ and $y_2$, of two 
random variables, $x_1$ and $x_2$, can be found from

\begin{equation}
f_{Y_1Y_2}(y_1,y_2) = f_{X_1X_2}(x_1,x_2)|J(x_1,x_2)|^{-1}
\label{eqn:joint}
\end{equation}

\noindent where $f_{X_1X_2}(x_1,x_2) = f_{X_1}(x_1)f_{X_2}(x_2)$ is the joint probability 
density of the independent RVs $x_1$ and $x_2$, and the matrix $J$ is the Jacobian given by

\begin{equation}
J(x_1,x_2) = 
\begin{pmatrix}
\partial y_1/\partial x_1 & \partial y_1/\partial x_2 \\
\partial y_2/\partial x_1 & \partial y_2/\partial x_2 \\
\end{pmatrix}.
\end{equation}

\noindent From equations~\ref{eqn:fexp} and~\ref{eqn:joint}, the joint probability density 
of ${\rm I} = y_1=x_1+x_2$ and ${\rm Q} = y_2=x_1-x_2$ is

\begin{equation}
f_{\rm QI}(y_1,y_2) = {1\over{2\mu_1\mu_2}}
                  \exp\Biggl[-{y_1(\mu_1+\mu_2)\over {2\mu_1\mu_2}}\Biggr]
                  \exp\Biggl[{y_2(\mu_1-\mu_2)\over {2\mu_1\mu_2}}\Biggr].
\label{eqn:fQI}
\end{equation}

\noindent The cumulative distribution of the fractional linear polarization, $F_{\rm m}(z)$, 
is the probability that ${{\rm |Q/I|}\le z}$ (MS1).

\begin{equation}
F_{\rm m}(z) = P\{{\rm |Q/I|}\le z\} = P\{-z\le{\rm Q/I}\le z\}
       = F_{\rm Q/I}(z) - F_{\rm Q/I}(-z)
\end{equation}

\begin{equation}
F_{\rm m}(z)=\int_0^\infty\int_{-\infty}^{y_1z}f_{\rm QI}(y_1,y_2)dy_2dy_1
         -\int_0^\infty\int_{-\infty}^{-y_1z}f_{\rm QI}(y_1,y_2)dy_2dy_1
\end{equation}

\noindent The cumulative distribution of the fractional linear polarization for the specific
case of exponentially distributed mode intensities is found by completing the above integrals
using the joint probability density of I and Q from equation~\ref{eqn:fQI}.

\begin{equation}
F_{\rm m}(z) = {(1-m^2)z\over{1-m^2z^2}}
\end{equation}

\noindent The distribution of fractional polarization given by equation~\ref{eqn:mdist} 
is then found by taking the derivative of the cumulative probability density.

\begin{equation}
f_{\rm m}(z) = {dF_{\rm m}\over {dz}}
\end{equation}

\acknowledgments{The National Radio Astronomy Observatory is a facility of the National 
Science Foundation operated under cooperative agreement by Associated Universities, Inc.}

\end{document}